\documentclass[twoside,draft]{article}
\usepackage{latexsym,e-journal-new}

\usepackage[utf8]{inputenc}       
\usepackage[english]{babel} 

\usepackage{amssymb,amsfonts}       

\usepackage{dsfont}
\usepackage{bbm}

\usepackage[perpage,symbol*]{footmisc}
\usepackage[final]{graphicx}                             
\usepackage{pstricks}

\usepackage{cite}

\usepackage[varg]{txfonts}

\begin{document}

\newcommand{\rum}{\rule{0.5pt}{0pt}}
\newcommand{\rub}{\rule{1pt}{0pt}}
\newcommand{\rim}{\rule{0.3pt}{0pt}}
\newcommand{\numtimes}{\mbox{\raisebox{1.5pt}{${\scriptscriptstyle \rum\times}$}}}
\newcommand{\numtimess}{\mbox{\raisebox{1.0pt}{${\scriptscriptstyle \rum\times}$}}}
\newcommand{\Boldsq}{\vbox{\hrule height 0.7pt
\hbox{\vrule width 0.7pt \phantom{\footnotesize T}%
\vrule width 0.7pt}\hrule height 0.7pt}}
\newcommand{\two}{$\raise.5ex\hbox{$\scriptstyle 1$}\kern-.1em/
\kern-.15em\lower.25ex\hbox{$\scriptstyle 2$}$}

\renewcommand{\refname}{References}
\renewcommand{\tablename}{\small Table}
\renewcommand{\figurename}{\small Fig.}
\renewcommand{\contentsname}{Contents}

\twocolumn[%
\begin{center}
\renewcommand{\baselinestretch}{0.93}
{\Large\bfseries  Lee Smolin Five Great Problems and Their Solution\\ without Ontological Hypotheses

}\par
\renewcommand{\baselinestretch}{1.0}
\bigskip
Gunn Quznetsov
\\ 
{\footnotesize  Chelyabinsk State University, Chelyabinsk, Ural, Russia.\rule{0pt}{12pt}
E-mail: gunn@mail.ru, quznets@yahoo.com\\

}\par
\medskip
{\small\parbox{11cm}{%
Solutions of Lee Smolin Five Great Problems from his book {\it The Trouble with Physics: 
the Rise of String Theory, the Fall of a Science, and What Comes Next} are described. This solutions is obtained only 
from the properties of probability without any ontological hypotheses.
}}\smallskip
\end{center}]{%

\setcounter{section}{0}
\setcounter{equation}{0}
\setcounter{figure}{0}
\setcounter{table}{0}
\setcounter{page}{1}

\markboth{Gunn Quznetsov. Lee Smolin Five Great Problems}{\thepage}
\markright{Gunn Quznetsov, Lee Smolin Five Great Problems}

\section*{Introduction}
\markright{Gunn Quznetsov, Lee Smolin Five Great Problems}

In his book \cite{S} Lee Smolin, professor of Perimeter Institute, Canada, has formulated the following five problems 
which he named Great Problems:

\begin{quote}
{\it Problem 1: Combine general relativity and qua-\linebreak ntum theory into a single theory that claim to be the complete 
theory of nature.}
\end{quote}

\begin{quote}
{\it Problem 2: Resolve the problems in the foundations of quantum mechanics, either by making sense of the theory as 
it stands or by inventing a new theory that does make sense. \dots\ }
\end{quote}

\begin{quote}
{\it Problem 3: Determine whether or not the various particles and forces can be unified in a theory that explain them 
all as manifestations of a single, fundamental entity. \dots\ }
\end{quote}

\begin{quote}
{\it Problem 4: Explain how the values of of the free constants in the standard model of particle phy-\linebreak sics are chosen 
in nature. \dots\ }
\end{quote}

\begin{quote}
{\it Problem 5: Explain dark matter and dark energy. Or if they don't exist, determine how and why gravity is 
modified on large scales. \dots\ }
\end{quote}

\section*{Solution}
\markright{Gunn Quznetsov, Lee Smolin Five Great Problems}

Let us consider the free Dirac Lagrangian:

\begin{equation}
\mathcal{L}:=\psi ^{\dagger }\left( \beta ^{\left[ k\right] }\partial
_k+m\gamma ^{\left[ 0\right] }\right) \psi \mbox{.}  \label{1}
\end{equation}

Here\footnote{$0_2:=\left[ 
\begin{array}{cc}
0 & 0 \\ 
0 & 0
\end{array}
\right] ,1_2:=\left[ 
\begin{array}{cc}
1 & 0 \\ 
0 & 1
\end{array}
\right] ,$
$\beta ^{\left[ 0\right] }:=-1_4:=-\left[ 
\begin{array}{cc}
1_2 & 0_2 \\ 
0_2 & 1_2
\end{array}
\right] $, $k\in \left\{ 0,1,2,3\right\} $
, $\nu \in \left\{1,2,3\right\} $.
}
\[
\beta ^{\left[ \nu \right] }:=\left[ 
\begin{array}{cc}
\sigma _\nu & 0_2 \\ 
0_2 & -\sigma _\nu
\end{array}
\right] \quad \mbox{,  } \gamma ^{\left[ 0\right]
}:=\left[ 
\begin{array}{cc}
0_2 & 1_2 \\ 
1_2 & 0_2
\end{array}
\right] 
\]
where $\sigma _1$, $\sigma _2$, $\sigma _3$ are the Pauli matrices.

Such Lagrangian is not invariant \cite{Kane} under the SU(2)\linebreak transformation with the parameter~$\alpha $:
\begin{eqnarray*}
&&\psi ^{\dagger }U^{\dagger }(\alpha )\left( \beta ^{\left[ k\right]
}\partial _k+m_1\gamma ^{\left[ 0\right] }\right) U(\alpha )\psi = \\
&&=\psi ^{\dagger }\left( \beta ^{\left[ k\right] }\partial _k+\left( m\cos
\alpha \right) \gamma ^{\left[ 0\right] }\right) \psi \mbox{,}
\end{eqnarray*}
the mass member is changed under this transformation.

Matrices $\beta ^{\left[ \nu \right] }$ and $\gamma ^{\left[ 0\right] }$ are
anticommutative. But it turns out that there exists a fifth matrix $\beta ^{[4]} $ anticommuting with all these four 
matrices: 
\[
\beta ^{\left[ 4\right] }:=\mathrm{i}\left[ 
\begin{array}{cc}
0_2 & 1_2 \\ 
-1_2 & 0_2
\end{array}
\right] \mbox{.}
\]

And the term with this matrix should be added to this Lagrangian mass term:
\[
\underline{\mathcal{L}}:=\psi ^{\dagger }\left( \beta ^{\left[ k\right]
}\partial _k+m_1\gamma ^{\left[ 0\right] }+m_2\beta ^{\left[ 4\right]
}\right) \psi 
\]
where $\sqrt{m_1^2+m_2^2}=m$.

Let $U(\alpha )$ be any SU(2)-matrix with parameter $\alpha $ and let $\mathbf{U%
}$ be the space in which $U(\alpha )$ acts. In such case $U(\alpha )$
divides the space $\mathbf{U}$ into two orthogonal subspaces $\mathbf{U}_o$
and $\mathbf{U}_x$ such that for every element $\psi $ of $\mathbf{U}$ there
exists an element $\psi _o$ of $\mathbf{U}_o$ and an element $\psi _x$ of $%
\mathbf{U}_x$ which fulfills the following conditions \cite{PTGT, PTJ}:

1. 
\[
\psi _o+\psi _x=\psi \mbox{,}
\]

2. 
\begin{eqnarray}
&&\psi _o^{\dagger }U^{\dagger }(\alpha )\left( \beta ^{\left[ k\right]
}\partial _k+m_1\gamma ^{\left[ 0\right] }+m_2\beta ^{\left[ 4\right]
}\right) U(\alpha )\psi _o= \nonumber  \\
&=&\psi _o^{\dagger }(\beta ^{\left[ k\right] }\partial _k+\left( m_1\cos
\alpha -m_2\sin \alpha \right) \gamma ^{\left[ 0\right] }+ \label{Lo} \\
&&+\left( m_2\cos \alpha +m_1\sin \alpha \right) \beta ^{\left[ 4\right]
})\psi _o\mbox{,} \nonumber
\end{eqnarray}

3. 
\begin{eqnarray}
&&\psi _x^{\dagger }U^{\dagger }(\alpha )\left( \beta ^{\left[ k\right]
}\partial _k+m_1\gamma ^{\left[ 0\right] }+m_2\beta ^{\left[ 4\right]
}\right) U(\alpha )\psi _x= \nonumber \\
&=&\psi _x^{\dagger }(\beta ^{\left[ k\right] }\partial _k+\left( m_1\cos
\alpha +m_2\sin \alpha \right) \gamma ^{\left[ 0\right] }+ \label{Lx} \\
&&+\left( m_2\cos \alpha -m_1\sin \alpha \right) \beta ^{\left[ 4\right]
})\psi _x \mbox{.} \nonumber
\end{eqnarray}
In either case, $m$ does not change.

I call these five ($\beta :=\left\{ \beta ^{\left[ \nu \right] },\beta ^{\left[ 4\right]
},\gamma ^{\left[ 0\right] }\right\} $) anticommuting matrices \textit{Clifford pentad}. Any
sixth matrix does not anticommute with all these five.

There exist only six Clifford pentads (for instance, \cite{MD, LFTP}): 

I call one of them (the pentad $\beta $) 
{\it the light pentad}, three ($\zeta $, $\eta $, $\theta $) --- {\it the chromatic pentads}, and two 
( $\underline{\Delta }$, $\underline{\Gamma }$) --- {\it the gustatory pentads}. 

The light pentad contains three diagonal matrices ($\beta ^{\left[ \nu \right] }$) corresponding to the coordinates 
of 3-dimensional space, and two antidiagonal 
matrices ($\beta ^{\left[ 4\right] }$, $\gamma ^{\left[ 0\right] }$)
relevant to mass terms (\ref{Lo},\ref{Lx}) --- one for the lepton state and the other for the neutrino state of this lepton.

Each chromatic pentad also contains three diagonal matrices corresponding to three coordinates and 
two antidiagonal mass matrices - one for top quark state and the other --- for bottom quark state.

Each gustatory pentad contains a single diagonal coordinate matrix 
and two pairs of antidiagonal mass matrices \cite{LFTP} --- these pentads are not needed yet.

Let\footnote{$\mathrm{c} = 299792458$.} $\left\langle \rho _{A}\mathrm{c} ,j_{A,\nu }\right\rangle $ be a
1+3-vector of probability density of a pointlike event $A$.

For any $A$ the set of four equations with an unknown complex $4\times 1$ matrix
function~$\varphi (x_k)$
\[
\left\{ 
\begin{array}{c}
\rho _A=\varphi ^{\dagger }\varphi \mbox{,} \\ 
\displaystyle \frac{j_{A,\nu }}{\mathrm{c}}=-\varphi ^{\dagger }\beta ^{\left[ \nu \right]
}\varphi 
\end{array}
\right| 
\]
has solution \cite{PTGT}.

If\footnote{$\mathrm{h}:=$ $6.6260755\cdot 10^{-34}$} $\rho _{\mathtt{A}}\left( x_k\right) =0$ for 
all $x_k$ such that $\left| x_k \right| >\left( \pi \mathrm{c}/\mathrm{h}%
\right) $  then $\varphi $ obeys the following equation~\cite{PP}:
\begin{eqnarray*}
&&\rule{-0.7cm}{0pt}\biggl ( \biggr. 
\!\!{-}\!\left( \partial _0{+}\mathrm{i}\Theta _0{+}\mathrm{i}\Upsilon
_0\gamma ^{\left[ 5\right] }\right) {+}\beta ^{\left[ \nu \right] }\left(
\partial _\nu {+}\mathrm{i}\Theta _\nu {+}\mathrm{i}\Upsilon _\nu \gamma
^{\left[ 5\right] }\right)+  \\
&&{+}2\left( \mathrm{i}M_0\gamma ^{\left[ 0\right] }{+}\mathrm{i}M_4\beta
^{\left[ 4\right] }\right) \biggl. \biggr ) \varphi {+} \\
&&\rule{-1cm}{0pt}{+}\biggl ( \biggr. 
\!\!{-}\!\left( \partial _0{+}\mathrm{i}\Theta _0{+}\mathrm{i}\Upsilon
_0\gamma ^{\left[ 5\right] }\right) {-}\zeta ^{[\nu ]}\left( \partial _\nu {+%
}\mathrm{i}\Theta _\nu {+}\mathrm{i}\Upsilon _\nu \gamma ^{\left[ 5\right]
}\right)+  \\
&&{+}2\left( {-}\mathrm{i}M_{\zeta ,0}\gamma _\zeta ^{[0]}{+}\mathrm{i}%
M_{\zeta ,4}\zeta ^{[4]}\right) \biggl. \biggr ) \varphi {+} \\
&&\rule{-1cm}{0pt}{+}\biggl ( \biggr.
\!\!\left( \partial _0{+}\mathrm{i}\Theta _0{+}\mathrm{i}\Upsilon _0\gamma
^{\left[ 5\right] }\right) {-}\eta ^{[\nu ]}\left( \partial _\nu {+}\mathrm{i%
}\Theta _\nu {+}\mathrm{i}\Upsilon _\nu \gamma ^{\left[ 5\right] }\right)+  \\
&&{+}2\left( {-}\mathrm{i}M_{\eta ,0}\gamma _\eta ^{[0]}{-}\mathrm{i}M_{\eta
,4}\eta ^{[4]}\right) \biggl. \biggr ) \varphi {+} \\
&&\rule{-1cm}{0pt}{+}\biggl ( \biggr. 
\!\!{-}\!\left( \partial _0{+}\mathrm{i}\Theta _0{+}\mathrm{i}\Upsilon
_0\gamma ^{\left[ 5\right] }\right) {-}\theta ^{[\nu ]}\left( \partial _\nu {%
+}\mathrm{i}\Theta _\nu {+}\mathrm{i}\Upsilon _\nu \gamma ^{\left[ 5\right]
}\right)+  \\
&&{+}2\left( \mathrm{i}M_{\theta ,0}\gamma _\theta ^{[0]}{+}\mathrm{i}%
M_{\theta ,4}\theta ^{[4]}\right) \biggl. \biggr ) \varphi=  \\
&=&0
\end{eqnarray*}
with real\\ $\Theta _k\left( x_k\right) $, $\Upsilon _k\left( 
x_k\right) $, $M_0\left( x_k\right) $, $M_4\left( 
x_k\right) $, $M_{\zeta ,0}\left( x_k\right) $,\\ $%
M_{\zeta ,4}\left( x_k\right) $, $M_{\eta ,0}\left( x_k%
\right) $, $M_{\eta ,4}\left( x_k\right) $, $M_{\theta ,0}\left( 
x_k\right) $, $M_{\theta ,4}\left( x_k\right) $\\ and with
\[
\gamma ^{\left[ 5\right] }:=\left[ 
\begin{array}{cc}
1_2 & 0_2 \\ 
0_2 & -1_2
\end{array}
\right] \mbox{.} 
\]

The first summand of this equation contains elements of the light pentad only. And the rest summands contain elements 
of the chromatic pentads only.

This equation can be rewritten in the following way:
\begin{eqnarray}
&&\beta ^{\left[ k\right] }\left( -\mathrm{i}\partial _k+\Theta _k+\Upsilon
_k\gamma ^{\left[ 5\right] }\right) \varphi +  \label{ham0} \\
&&+(M_0\gamma ^{\left[ 0\right] }+M_4\beta ^{\left[ 4\right] }-  \nonumber \\
&&-M_{\zeta ,0}\gamma _\zeta ^{[0]}+M_{\zeta ,4}\zeta ^{[4]}-  \nonumber \\
&&-M_{\eta ,0}\gamma _\eta ^{[0]}-M_{\eta ,4}\eta ^{[4]}+  \nonumber \\
&&+M_{\theta ,0}\gamma _\theta ^{[0]}+M_{\theta ,4}\theta ^{[4]})\varphi = 
\nonumber \\
&=&0  \nonumber
\end{eqnarray}
because 
\[
\zeta ^{[\nu]}{+}\eta ^{[\nu]}{+}\theta ^{[\nu]}={-}\beta ^{\left[ \nu\right] }\mbox{.} 
\]

This equation is a generalization of the Dirac's equation with gauge fields  $\Theta _k\left( x_k\right) $ and $\Upsilon _k\left( 
x_k\right) $ and with eight mass members. The mass members with elements of the light pentad ( $M_0 $ and $M_4 $) 
conform to neutrino and its lepton states. And six mass members with elements of the chromatic pentads conform to three 
pairs (up and down) of chromatic states (red, green, blue).


Let this equation not contains the chromatic mass nembers:

\begin{equation}
\ \left( \beta ^{\left[ k\right] }\left( -\mathrm{i}\partial _k+\Theta
_k+\Upsilon _k\gamma ^{\left[ 5\right] }\right) +M_0\gamma ^{\left[ 0\right]
}+M_4\beta ^{\left[ 4\right] }\right) \varphi =0 \mbox{.}  \label{Lep}
\end{equation}

If function $\varphi $ is a solution of this equation then $\varphi $
represents the sum of functions $\varphi _{n,s}$ which satisfy the following
conditions \cite[62--71]{PTGT}:

$n$ and $s$ are integers;

each of these functions obeys its equation of the 
following form:

\begin{equation}
\left( \beta ^{\left[ k\right] }\left( \mathrm{i}\partial _k-\Theta
_0-\Upsilon _0\gamma ^{\left[ 5\right] }\right) -\frac{\mathrm{h}}{\mathrm{c}%
}\left( \gamma ^{\left[ 0\right] }n+\beta ^{\left[ 4\right] }s\right)
\right) \varphi _{n,s}=0 \mbox{;} \label{Lep1}
\end{equation}

for each point $x_k$ of space-time: or this point is empty (for all $n$ and $%
s$: $\varphi _{n,s}\left( x_k\right) =0$), 
or in this point is placed a single function
(for $x_k$ there exist integers $n_0$ and $s_0$ such that $%
\varphi _{n_0,s_0}\left( x_k\right) \neq 0$ and if $n\neq n_0$ and/or $s\neq
s_0$ then $\varphi _{n,s}\left( x_k\right) =0$).

In this case if $m:=\sqrt{n^2+s^2}$ then $m$ is a natural number. But under
the SU(2)-transformation with parameter $\alpha $ (\ref{Lo}, \ref{Lx}):\\ $m\rightarrow 
 \left( \left( n\cos \alpha -s\sin \alpha \right) ^2+\left( s\cos \alpha
+n\sin \alpha \right) ^2\right) ^{0.5} $,
\\ $\left( n\cos \alpha -s\sin \alpha \right) $ and $\left(
s\cos \alpha +n\sin \alpha \right) $ must be integers\linebreak too.
But it is impossible.

But for arbitrarily high accuracy in distant areas of the natural scale there
exist such numbers $m$ that for any $\alpha $ some natural numbers $n^{\prime }$ and $s^{\prime }$
exist which obey the following conditions: $n^{\prime }\approx \left( n\cos \alpha -s\sin
\alpha \right) $ and $s^{\prime }\approx \left( s\cos \alpha +n\sin \alpha
\right) $. These numbers $m$ are separated by long intervals and
determine the mass spectrum of the generations of elementary particles. Apparently, this is the way to solve Problem 4 
because the masses are one of the most important constants of particle physics.

The Dirac's equation for leptons with gauge members\linebreak
which are similar to electroweak fields is obtained \cite[p.333--336]{PTJ} from equations (\ref{Lep}, \ref{Lep1}). 
Such equation is invariant under electroweak transformations. And here the fields $W$ and $Z$ obey the Klein-Gordon
type equation with nonzero mass.


If the equation (\ref{ham0}) does not contain lepton's and neutrino's mass terms then the Dirac's equation with gauge
members which are similar to eight gluon's fields is obtained. And oscillations
of the chromatic states of this equation bend space-time. This
bend gives rise to the effects of redshift, confinement and
asymptotic freedom, and Newtonian gravity turns out to be a
continuation of subnucleonic forces \cite{PP}. And it turns out that these oscillations bend
space-time so that at large distances the space expands with acceleration according to
Hubble's law \cite{PP3}. And these oscillations bend space-time so that here appears the discrepancy between the 
quantity of the luminous matter in the space structures and the traditional picture of gravitational interaction 
of stars in these structures. Such curvature explains this discrepancy without the Dark Matter hypothesis \cite{DM} 
(Problem 5).

Consequently, the theory of gravitation is a continuation of quantum theory (Problem 1 and Problem 3). 

Thus, concepts and statements of Quantum Theory are concepts and statements of the probability of pointlike 
events and their ensembles.

Elementary physical particle in vacuum behaves like the-\linebreak se probabilities. For example, according to doubleslit
experiment \cite{Arx}, if a partition with two slits is placed between a source of elementary particles 
and a detecting screen in vacuum then interference occurs. But if this system will be put in a cloud chamber, then 
 trajectory of a particle will be clearly marked with drops of condensate and any interference will disappear. It seems that 
a physical particle exists only in the instants of time when some events happen to it. And in the other instants of time 
the particle does not exist, but the probability of some event to happen to this particle remains.

Thus, if no event occurs between an event of creation of a particle and an event of detection of it, then the particle does 
not exist in this period of time. There exists only the probability of detection of this particle at some point.
But this probability, as we have seen, obeys the equations of quantum theory and we get the interference. But in a cloud 
chamber events of condensation form a chain meaning the trajectory of this particle. In this case the interference 
disappears. But this trajectory is not continuous --- each point of this line has an adjacent point. And the effect of 
movement of this particle arises from the fact that a wave of probability propagates between these points.

Consequently, the elementary physical particle represents an ensemble of pointlike events associated with probabilities.
And charge, mass, energy, momentum, spins, etc. represent parameters of distribution of these probabilities.
It explains all paradoxes of quantum physics. Schr\"{o}dinger's cat lives easily without any superposition
of states until the microevent awaited by everyone occurs. And the wave function disappears without 
any collapse in the moment when event probability disappears after the event occurs.

Hence, entanglement concerns not particles but probabilities. That is when the event of the measuring of spin of Alice's 
electron occurs then probability for these entangled electrons is changed instantly in the whole space. Therefore, 
nonlocality acts for probabilities, not for particles. But probabilities can not transmit any information (Problem 2).

\section*{Conclusion}
\markright{Gunn Quznetsov, Lee Smolin Five Great Problems}

Therefore, Lee Smolin's Five Great Problems do have solution only using the properties of probabilities. These solutions 
do not require any dubious ontological hypotheses such as superstrings, spin networks, etc.

\begin{flushright}\footnotesize
Submitted on December 15, 2010 / Accepted on December 16, 2010
\end{flushright}

}
\vspace*{-6pt}
\centerline{\rule{72pt}{0.4pt}}


\begin{thebibliography}{99}\footnotesize

\bibitem{S} Smolin L. The trouble with physics: the rise of string theory, the fall of a science, and what comes next.
Houghton Mifflin, Boston, 2006.
\bibitem{Kane} Kane G. Modern Elementary Particle Physics. Addison-Wesley Publ. Comp., 1993, 93.
\bibitem{PTGT} Quznetsov, G. Probabilistic Treatment of Gauge Theories, in series {\it Contemporary Fundamental Physics}, 
ed. Dvoeglazov V., Nova Sci. Publ., N.Y., 2007.

\bibitem{PTJ} Quznetsov G. It is not Higgs, \emph{Prespacetime Journal}, 2010, v.1, 314--343


\bibitem{MD} Madelung E. Die Mathematischen Hilfsmittel des Physikers. Springer Verlag, Berlin, Gottingen, 
Heidelberg, 1957, 12.
\bibitem{LFTP} Quznetsov G. Logical Foundation of Theoretical Physics. Nova Sci. Publ., N.Y., 2006.

\bibitem{PP3} Quznetsov G. A. Oscillations of the Chromatic States and Accelerated Expansion of the Universe. 
\emph{Progress in Physics},  2010, v.2, 64--65.

\bibitem{DM} Quznetsov G. Dark Matter and Dark Energy are Mirage. {\it Prespacetime Journal}, October 2010, v.1, Issue 8, 1241-1248,
arXiv: 1004.4496 2

\bibitem{Arx} Quznetsov G. Double-Slit Experiment and Quantum Theory Event-Probability Interpretation, arXiv: 1002.3425.

\bibitem{PP} Quznetsov G. A. 4X1-Marix Functions and Dirac's Equation. \emph{Progress in Physics}, 2009, v.2, 96--106.



\end{thebibliography}
\end{document}